\documentclass[runningheads]{llncs}

\usepackage[T1]{fontenc}
\usepackage[table]{xcolor}
\usepackage{multirow}
\usepackage{booktabs}
\usepackage{array}
\usepackage{tabularx}
\usepackage{colortbl}
\usepackage{algorithm}
\usepackage{algpseudocode}
\usepackage{enumitem}
\usepackage{amsmath}
\usepackage{tikz}
\usepackage{wrapfig}
\usepackage{float}
\usepackage{hyperref}

\newcommand{\sscode}[1]{\textsf{#1}}
\newcommand{\code}[1]{\texttt{#1}}

\begin{document}

\title{Graph-Based Discovery of Mathematical Software Communities and Publication-to-Community Prediction}
\titlerunning{Graph-Based Discovery of Mathematical Software Communities}

\author{Maxence Azzouz-Thuderoz \and
Yuni Susanti \and
Moritz Schubotz}
\authorrunning{M. Azzouz-Thuderoz et al.}

\institute{FIZ Karlsruhe, Berlin, Germany\\
\email{\{maxence.azzouz, yuni.susanti, moritz.schubotz\}@fiz-karlsruhe.de}}

\maketitle

\begin{abstract}

Research software forms distinct co-usage communities that span traditional disciplinary boundaries, yet the structure of these communities remains largely unexplored. We present a graph-based framework for discovering mathematical software communities and predicting their association with research publications. 
We construct a software co-usage network from publication--software relationships using a curated \textsf{swMATH} dataset and subsequently apply community detection method, revealing a heterogeneous landscape of mathematical software communities. We formulate \textit{publication-to-community} mapping as a multi-label classification task and further investigate whether community membership can be predicted from lightweight scholarly metadata. Specifically, we compare two feature representations of scientific publications: Mathematics Subject Classification (MSC) and title-based embeddings. Across a range of models, structured MSC representation consistently provides a stronger precision--recall trade-off, demonstrating that structured domain metadata captures software-community structure more effectively than compressed title-only semantics in this setting. This work highlight the continuing value of structured scholarly metadata for large-scale research software discovery, classification and recommendation.

\end{abstract}

\section{Introduction}

Research software has become an essential component of modern mathematical practice, enabling computational methods and reproducible workflows. As mathematical research increasingly depends on specialized software ecosystems, systematic approaches for discovering and understanding software usage patterns across the literature remain limited~\cite{IstrateFYM24,DruskatHBK24}.
Curated resources such as \sscode{swMATH}~\cite{10.1007/978-3-662-44199-2_103} address this need by cataloging more than 32,000 mathematical software packages and over 628,000 publication--software links.
These links are established through editorial curation, supported by software-mention suggestions during the editorial workflow, resulting in a high-quality knowledge base that captures the relationship between research publications and software.

Existing approaches to linking publications and software aim to identify explicit software mentions in full text using named entity recognition (NER) and large language models (LLMs)~\cite{OttoUD24,NguyenThiNDL24,NguyenXuanTD24}. While effective, these methods depend on access to full-text documents, which may be unavailable because of publisher restrictions, and inherently limited to software that is explicitly mentioned. 
In this work, we investigate a problem of software \emph{community} prediction, focusing on the prediction of groups of software packages that exhibit consistent co-usage patterns across literature. Unlike approaches that recover individual software mentions, community-level prediction captures broader patterns of software adoption and can identify related tools even when they are not explicitly referenced in a publication. We focus on the following research questions:

\begin{enumerate}[label=\textbf{RQ\arabic*:}, leftmargin=1.5cm]
    \item How is the mathematical research software ecosystem organized into communities based on patterns of software usage in the literature?
    \item Can publication-level scholarly bibliographic metadata (e.g., title, subject classification i.e., MSC codes) support reliable prediction of software community membership?
\end{enumerate}

To answer these questions, we first construct a weighted software co-usage graph from publication--software links in the \sscode{swMATH} database and identify software communities using a standard community detection algorithm. We then formulate \textit{publication-to-community} mapping as a multi-label classification task by assigning each publication the communities of the software packages it cites. We further investigate whether these community labels can be predicted from lightweight publication-level scholarly metadata by comparing two complementary representations: expert-curated Mathematics Subject Classification (MSC) codes and neural embeddings derived from publication titles.
This comparison reflects a realistic large-scale scenario in which only standard scholarly metadata are readily available. Although graph-based community detection has been widely applied to citation, collaboration, and other scholarly networks, comparatively little attention has been devoted to uncovering the structure of mathematical software ecosystems through publication--software relationships. By modeling software co-usage as a network, our approach provides a data-driven characterization of software communities and establishes a foundation for predicting publication-to-community associations from lightweight scholarly metadata.

\paragraph{Contributions.}
Our main contributions are threefold. First, we present a graph-based framework for discovering mathematical software communities from publication--software relationships. Second, we formulate publication-to-community mapping as a multi-label classification task and systematically compare two lightweight scholarly metadata representations for predicting software community membership. Third, we show that the structured subject classification metadata consistently yield better precision--recall trade-off than title-based embeddings, highlighting the continuing value of structured scholarly metadata for large-scale software discovery, recommendation, and research knowledge organization.
We publicly release all data and resources to support future research.\footnote{https://github.com/swMATH/Paper2Community}

\section{Related Work}

Recent research highlights the growing importance of software in scholarly communication. A substantial body of work has focused on identifying and characterizing software references in scientific publications. \cite{IstrateFYM24} proposed a framework for classifying software citation intent, demonstrating that language models such as BERT and GPT can achieve over 80\% accuracy~\cite{10.1145/3677389.3702514}. \cite{StankovskiG24} introduced a method to automatically identifies the primary GitHub repository associated with a research paper by analyzing the context of code mentions in text. Other studies have leveraged generative LLMs and retrieval-augmented generation (RAG) for software named entity recognition~\cite{OttoUD24,NguyenThiNDL24,NguyenXuanTD24}. These efforts contribute to broader initiatives promoting research transparency and reproducibility~\cite{DruskatHBK24,KrugerKD24}, including the integration of software mentions into knowledge graphs~\cite{CiuciuKissG24}. More recently, \textit{SemRepo}~\cite{semrepo} introduced a large-scale RDF knowledge graph that links GitHub repositories with scientific publications and the broader scholarly ecosystem, allowing large-scale analysis of research software sustainability.

Parallel research has investigated the social and organizational dynamics of open-source software communities. \cite{ALMARIMI2020106201} introduced \textit{csDetector}, a framework for detecting ``community smells'', which was later extended with socio-technical and sentiment analysis in~\cite{https://doi.org/10.1002/smr.2505}. Complementing this line of work, \cite{10.1145/3487351.3488278} demonstrated that geographical distribution and programming language significantly influence collaboration patterns on GitHub.
While existing research has primarily focused on software mention extraction, repository linking, knowledge graph construction, and the analysis of social dynamics within software communities, the automated classification of software into communities remains unexplored. Our work addresses this gap by (i) classifying publications into software co-usage communities rather than extracting individual software mentions, capturing ecosystem-level relationships; (ii) leveraging lightweight scholarly metadata (MSC codes, title) instead of full-text content, supporting scalable classification; and (iii) to the best of our knowledge, providing the first empirical comparison of domain taxonomies and neural embeddings for software community prediction.

\section{Methodology}

\subsection{Task Formulation}
We formulate software community prediction as a \emph{multi-label classification} problem of publication-to-community mapping. Given a research publication represented by lightweight scholarly metadata, i.e., its title and MSC codes, the objective is to predict the set of mathematical software communities associated with that publication. This formulation enables scalable community assignment without requiring full-text access or explicit software mention extraction.

\subsection{Dataset: \sscode{swMATH} Software Database}
The \sscode{swMATH} database is a curated catalog of mathematical research software, developed through \sscode{zbMATH Open}'s editorial process~\cite{10.1007/978-3-662-44199-2_103}.
It spans mathematical disciplines, from \textit{numerical analysis} and \textit{statistics} to \textit{topology} and \textit{number theory}, and includes approximately 32,000 software entries linked to over 628,000 citations, with an average of 19 citations per entry (range: 1--15,194).
Our experiment uses metadata obtained via the \sscode{zbMATH} public API\footnote{\url{https://api.zbmath.org/v1/}}.
Each software record is annotated with canonical references, supported operating systems, implementation languages, dependencies, and licensing information. The catalog covers both domain-specific tools (e.g., \textit{finite element solvers}) and general-purpose platforms (e.g., \textit{Python}). Software coverage in \sscode{swMATH} is highly uneven across MSC categories: the most represented classes are 68 (\textit{Computer science}; 17,915), 62 (\textit{Statistics}; 11,673), and 65 (\textit{Numerical analysis}; 11,258), while few categories (e.g., 36, 25, 95) have fewer than 3 instances each.

\subsection{Software Co-usage Network Construction}

To capture relationships among mathematical software, we first construct a software co-usage network from publication--software associations in the \textsf{swMATH} dataset and subsequently identify software communities through graph partitioning technique. We model the mathematical software ecosystem as an undirected, weighted \emph{co-usage} graph $G=(V,E)$, where $V$ denotes the set of software packages and the weight of an edge $E(s_i,s_j)$ corresponds to the number of publications that cite both $s_i$ and $s_j$. Publications act as external entities linking software packages through co-citation: when a publication cites multiple software packages, it induces weighted connections between every pair of co-cited software. Thus, the graph captures patterns of software co-usage in the literature rather than expert-defined relationships such as software dependencies or complementary functionality.

\begin{figure}
    \centering
    \includegraphics[width=1\linewidth]{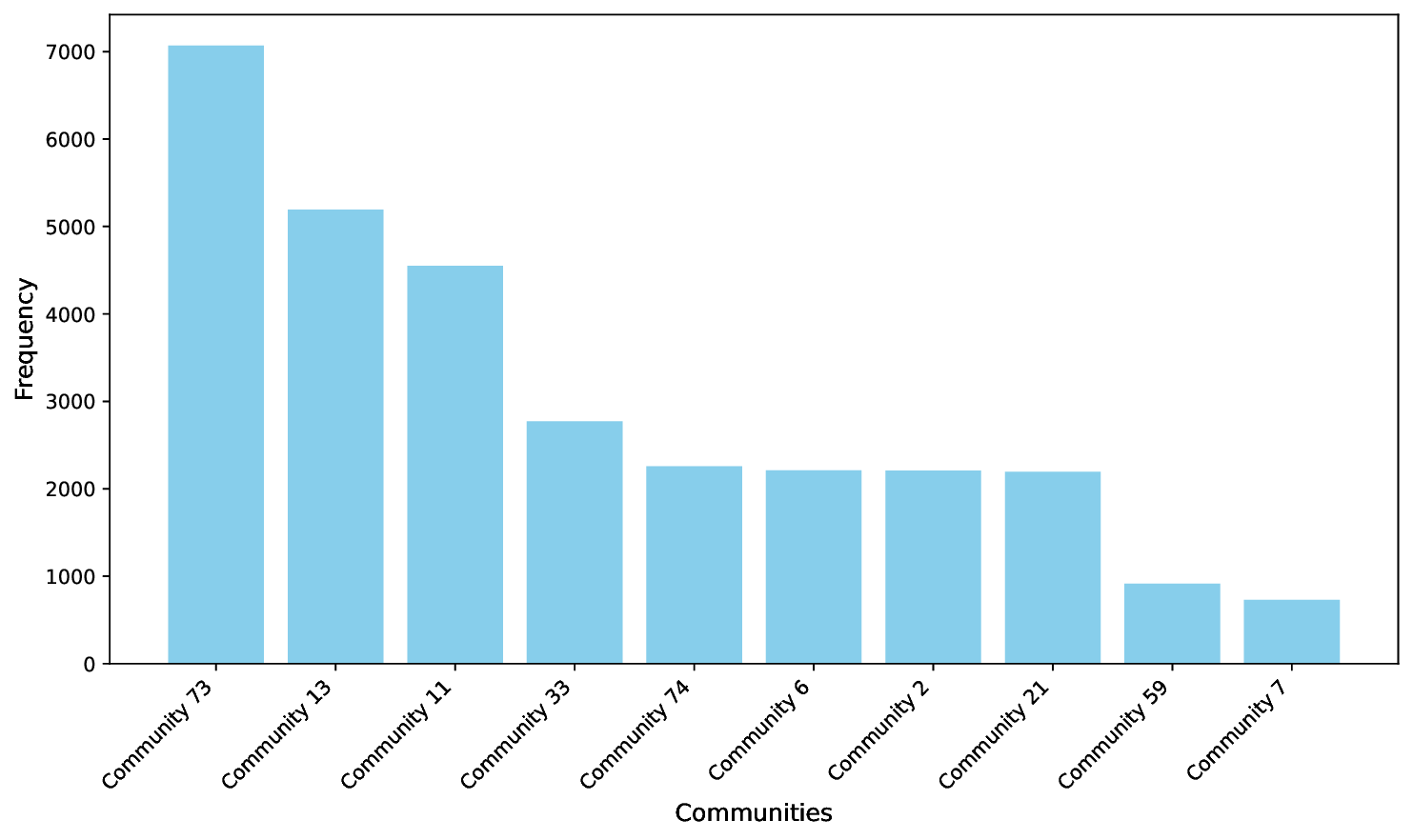}
    \caption{Top-10 mathematical software communities identified in \textsf{swMATH}, showing a heavy-tailed distribution.}
    \label{fig:comm}
\end{figure}

For each publication $p$, we extract the set $S_p$ of referenced software and create an undirected edge $(s_i,s_j)$ for every pair of distinct software tools $s_i,s_j\in S_p$. Edge weights correspond to the number of publications in which the pair is co-cited. To reduce noise, edges with weights below the threshold $\tau=2$ are removed, resulting in a weighted graph containing 32{,}987 software nodes and 388{,}963 edges.
Software communities are identified using the \textit{Louvain} community detection algorithm~\cite{Blondel_2008}, chosen for its scalability, modularity optimization, and automatic determination of the number of communities. This results into 197 communities with a modularity score of $Q=0.6852$, indicating a strong community structure, obtained with resolution values 1.0. A sensitivity analysis over resolution values in the range $[0.5,1.5]$ produced between 89 and 312 communities; we report the results for the default resolution of 1.0 as a representative trade-off between coarse and fine-grained partitions. The detailed hyperparameters is provided in our repository to facilitate reproducibility. 

Figure~\ref{fig:comm} shows the top 10 software communities identified in the swMATH dataset through this process. As visualized in the Figure~\ref{fig:comm}, the co-usage software network exhibits a heavy-tailed distribution, revealing a heterogeneous landscape of software communities with diverse disciplinary organization.

\subsection{Software Community Prediction}
\label{sec:features}
Each detected community represents a group of software tools that are frequently co-cited in the scientific literature. We map each publication to the communities associated with the software tools it references, formulating the community assignment as a multi-label classification task. Although a total 197 software communities are identified, only 170 are linked to at least one publication in the labeled corpus and are retained as prediction targets. The remaining 27 communities, which contain software not referenced by any publication in the corpus, are excluded from subsequent classification experiments.

For the multi-label classification, each publication is represented in one of two feature spaces: (i) a domain-specific representation derived from Mathematics Subject Classification (MSC) codes and (ii) a semantic representation based on title embeddings. Both representations are provided as input to the same set of multi-label classifiers, enabling a direct comparison of their effectiveness for predicting membership across the 170 software communities. The two representations are described in detail below.

\begin{figure}
  \centering
  \begin{tikzpicture}[
      scale=0.85, transform shape,
      every node/.style={circle, draw, minimum size=8mm, font=\tiny, align=center},
      ts/.style={fill=red!70},
      opt/.style={fill=cyan!70},
      stat/.style={fill=blue!50},
      ml/.style={fill=green!40},
      legendbox/.style={rectangle, draw=gray!50, fill=white, rounded corners, inner sep=4pt, font=\tiny}
  ]
  \node[opt] (esaddle)    at (4, 5)   {esaddle};
  \node[ts]  (tig)        at (0, 3.5) {Tigramite};
  \node[opt] (ltsa)       at (3, 3.5) {ltsa};
  \node[ts]  (fac)        at (1.5, 2.5) {factorstoch-\\vol};
  \node[ts]  (stfin)      at (-0.5, 1.5) {StFin-\\Metrics};
  \node[ts]  (art)        at (2.5, 1.5) {artfima};
  \node[opt] (orig)       at (4, 1.5)   {origami};
  \node[opt] (estima)     at (5.5, 2)   {ESTIMA};
  \node[opt] (vg)         at (6.5, 3)   {VG codes};
  \node[stat](funfits)    at (0.5, 0)   {FUNFITS};
  \node[ml]  (hsm)        at (1.5, -1)  {HSMClust};
  \node[ml]  (xpl)        at (3, -0.5)  {XploRe};
  \node[ml]  (mix)        at (4.5, -0.5){MixGHD};
  \node[opt] (chime)      at (6, 0.5)   {CHIME};
  \node[ml]  (clu)        at (3.5, -1.8){ClusterKDE};
  \draw (tig) -- (fac); \draw (tig) -- (ltsa);
  \draw (fac) -- (stfin); \draw (fac) -- (art); \draw (fac) -- (ltsa); \draw (fac) -- (funfits);
  \draw (ltsa) -- (esaddle); \draw (ltsa) -- (orig); \draw (ltsa) -- (estima); \draw (ltsa) -- (art);
  \draw (art) -- (orig); \draw (art) -- (xpl); \draw (art) -- (hsm); \draw (art) -- (funfits);
  \draw (orig) -- (estima); \draw (orig) -- (mix); \draw (orig) -- (chime);
  \draw (funfits) -- (hsm); \draw (funfits) -- (xpl);
  \draw (hsm) -- (xpl); \draw (hsm) -- (clu);
  \draw (mix) -- (clu); \draw (mix) -- (chime); \draw (mix) -- (xpl);
  \draw (estima) -- (vg); \draw (estima) -- (chime); \draw (vg) -- (chime);
  \draw (clu) -- (xpl);
  \node[legendbox, anchor=north] at (3, -2.8) {
      \begin{tabular}{@{}c@{\,}l@{\quad}c@{\,}l@{\quad}c@{\,}l@{\quad}c@{\,}l@{}}
          \tikz\node[ts, minimum size=3mm]{}; & Time Series &
          \tikz\node[opt, minimum size=3mm]{}; & Optimization &
          \tikz\node[stat, minimum size=3mm]{}; & Stat.\ Methods &
          \tikz\node[ml, minimum size=3mm]{}; & Machine Learning
      \end{tabular}
  };
  \end{tikzpicture}
  \caption{Subgraph of Community 73 (\textit{illustration}). Edges connect packages co-cited by at least $\tau=2$ publications; node colors represent manually-assigned specializations.}
  \label{fig:technical_domains_network}
\end{figure}
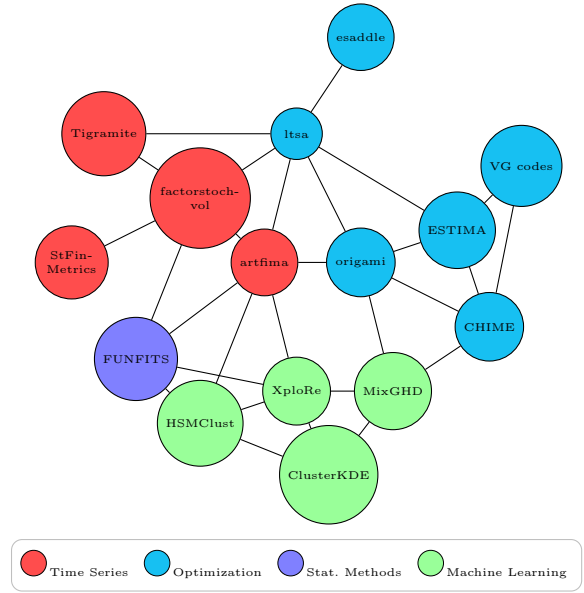

\paragraph{\textbf{MSC Representation.}}
The MSC representation is a 63-dimensional binary vector obtained by one-hot encoding the top-level MSC classes assigned to each paper. These expert-curated subject classifications provide a compact, domain-specific description of a publication and serve as input features for software community prediction.

The underlying software network exhibits a pronounced heavy-tailed structure. Among the ten largest communities (73, 13, 11, 33, 74, 6, 2, 21, 59, and 7; 30{,}098 tools in total), the three largest (73, 13, and 11) contain 16{,}813 tools (56\%), while Community~73 alone comprises 7{,}070 tools (21\% of the complete 32{,}987-node network). Community~13 is associated with 27.3\% of papers, followed by Community~73 (19.0\%) and Community~2 (15.6\%), with prevalence decreasing rapidly thereafter. Figure~\ref{fig:technical_domains_network} illustrates a curated subset of Community~73 with manually assigned technical specializations, highlighting clusters centered on \textit{optimization} and \textit{machine learning}.

\begin{table}
\centering
\caption{Research paper titles with color-coded words indicating software community group associations. Keywords are highlighted in bold.}
\label{tab:colored_titles}
\footnotesize
\setlength{\arrayrulewidth}{0.5pt}
\renewcommand{\arraystretch}{1.3}
\begin{tabularx}{0.86\linewidth}{|>{\raggedright\arraybackslash}X|}
\hline
\rowcolor{lightgray!30}
\multicolumn{1}{|c|}{\textbf{Research Paper Titles by Software Community Group}} \\
\hline
An \textcolor[RGB]{166,86,40}{\textbf{introduction}} to \textcolor[RGB]{166,86,40}{\textbf{analysis}} of financial \textcolor[RGB]{166,86,40}{\textbf{data}} with R. \\
\hline
\textcolor[RGB]{55,126,184}{\textbf{Deep}} \textcolor[RGB]{55,126,184}{\textbf{neural}} \textcolor[RGB]{55,126,184}{\textbf{networks,}} gradient-boosted trees, random forests: \textcolor[RGB]{166,86,40}{\textbf{statistical}} arbitrage on the \textcolor[RGB]{55,126,184}{\textbf{S\&P 500}} \\
\hline
\textcolor[RGB]{255,217,0}{\textbf{Exploring,}} handling, \textcolor[RGB]{255,217,0}{\textbf{imputing}} and evaluating \textcolor[RGB]{255,217,0}{\textbf{missing}} \textcolor[RGB]{166,86,40}{\textbf{data}} in \textcolor[RGB]{166,86,40}{\textbf{statistical}} analyses: a review of existing \textcolor[RGB]{255,217,0}{\textbf{approaches}} \\
\hline
\textcolor[RGB]{228,26,28}{\textbf{Computational}} \textcolor[RGB]{166,86,40}{\textbf{methods}} for \textcolor[RGB]{228,26,28}{\textbf{numerical}} \textcolor[RGB]{166,86,40}{\textbf{analysis}} with R \\
\hline
Fast likelihood calculation for \textcolor[RGB]{152,78,163}{\textbf{multivariate}} Gaussian \textcolor[RGB]{152,78,163}{\textbf{phylogenetic}} models with shifts \\
\hline
A \textcolor[RGB]{152,78,163}{\textbf{comparison}} of ancestral state reconstruction \textcolor[RGB]{166,86,40}{\textbf{methods}} for quantitative \textcolor[RGB]{152,78,163}{\textbf{characters}} \\
\hline
\end{tabularx}
\end{table}

\paragraph{\textbf{Title Embedding Representation.}}
The publication titles are encoded using the SPECTER model~\cite{Cohan2020SPECTERDR,specter2020cohan}, implemented through Sentence Transformers~\cite{reimers-2019-sentence-bert},
producing 768-dimensional embeddings optimized for scientific documents. To reduce computational cost, these embeddings are compressed to 128 dimensions using a symmetric autoencoder ($768 \rightarrow 512 \rightarrow 256 \rightarrow 128$) with ReLU activations, mean squared error loss, and early stopping. We selected 128 dimensions because reducing the embedding size to 64 dimensions increased reconstruction loss by approximately 18\%, whereas increasing it to 256 dimensions yielded less than a 2\% improvement while doubling memory requirements.

To isolate the effect of representation choice from input length, only paper titles are used for the embedding-based representation. Table~\ref{tab:colored_titles} presents representative titles together with manually highlighted domain-specific keywords.

\section{Evaluation}
We evaluate the proposed approach on the task of software community prediction, formulated as a multi-label classification problem. Given a research publication represented by its scholarly metadata, namely its title and MSC representations (see Section~\ref{sec:features} for details of the representation construction), the goal is to predict the set of mathematical software communities associated with the publication. This section describes the experimental setup and reports the evaluation result and discussion.

\subsection{Experimental Setup}
Experiments are conducted on a corpus of 164{,}552 papers, split randomly into training and test sets using an 80/20 ratio.
The split is not stratified, as standard stratification is not directly applicable to multi-label targets and no iterative multi-label stratification procedure was employed. 

To evaluate model robustness, experiments are conducted using 5-fold cross-validation on a randomly selected 10\% sample. The results are consistent across folds, with a recall variation of approximately $\pm$1.5\%. Performance is primarily assessed using micro-recall, which measures community coverage, and micro-F1 score. We prioritize recall for software discovery, as missing relevant communities is costlier than including candidates that can be filtered later~\cite{mustafa_comprehensive_2021,ahmed_sajid_novel_2023}.
To mitigate the severe class imbalance, we apply per-class oversampling of positive-label instances using \code{RandomOverSampler} on a flattened per-label representation of the training data. This increases the effective training set from 131{,}641 to 1{,}053{,}128 training instances by duplicating minority-label documents as required. 

We implement a PyTorch neural network~\cite{Ansel_PyTorch_2_Faster_2024} consisting of two fully connected layers with 512 hidden units and ReLU activation. Following the baseline setup of~\cite{tarekegn_deep_2024}, we compare our model against five traditional classifiers: Stochastic Gradient Descent (linear SVM with one-vs-rest classification), Passive-Aggressive (online margin-based learning), Multinomial Naive Bayes, Complement Naive Bayes (designed for imbalanced data), and L2-regularized Logistic Regression with one-vs-rest classification. All models are trained independently on each feature set. Implementation details are available in our repository (See Page~1).

\begin{table}[ht]
\centering
\scriptsize
\caption{Performance Metrics for MSC Codes and Title Embeddings Models}
\label{tab:combined_metrics}
\begin{tabular}{lcccccc|cccccc}
\toprule
\textbf{Metric} &
\multicolumn{6}{c|}{\textbf{MSC Codes Models}} &
\multicolumn{6}{c}{\textbf{Title Embeddings Models}} \\
\cmidrule(lr){2-7} \cmidrule(l){8-13}
&
\rotatebox{90}{Stochastic Gradient Descent} &
\rotatebox{90}{Passive Aggressive} &
\rotatebox{90}{Multinomial Naive Bayes} &
\rotatebox{90}{Complement Naive Bayes} &
\rotatebox{90}{Logistic Regression} &
\rotatebox{90}{Neural Network} &
\rotatebox{90}{Stochastic Gradient Descent} &
\rotatebox{90}{Passive Aggressive} &
\rotatebox{90}{Multinomial Naive Bayes} &
\rotatebox{90}{Complement Naive Bayes} &
\rotatebox{90}{Logistic Regression} &
\rotatebox{90}{Neural Network} \\
\midrule
Precision (Micro)     & \textbf{0.61} & 0.06 & 0.55 & 0.05 & 0.07 & 0.13 & 0.16 & 0.08 & \textbf{0.17} & 0.08 & 0.05 & 0.02 \\
Precision (Macro)     & 0.04 & 0.03 & 0.04 & 0.03 & 0.03 & \textbf{0.06} & 0.01 & 0.01 & 0.00 & 0.01 & 0.01 & 0.01 \\
Precision (Weighted)  & \textbf{0.60} & 0.52 & \textbf{0.60} & \textbf{0.60} & 0.50 & 0.59 & 0.15 & 0.15 & 0.14 & 0.15 & 0.15 & 0.15 \\
Precision (Samples)   & \textbf{0.64} & 0.14 & 0.60 & 0.11 & 0.14 & 0.32 & 0.16 & 0.08 & \textbf{0.17} & 0.08 & 0.05 & 0.03 \\
Recall (Micro)        & 0.64 & 0.65 & 0.65 & 0.67 & \textbf{0.79} & 0.66 & 0.47 & 0.92 & 0.77 & \textbf{0.97} & 0.48 & \textbf{0.97} \\
Recall (Macro)        & 0.04 & 0.12 & 0.05 & 0.18 & 0.21 & \textbf{0.61} & 0.02 & 0.08 & 0.03 & 0.07 & 0.07 & \textbf{0.68} \\
Recall (Weighted)     & 0.64 & 0.65 & 0.65 & 0.67 & \textbf{0.79} & 0.66 & 0.47 & 0.92 & 0.77 & \textbf{0.97} & 0.48 & \textbf{0.97} \\
Recall (Samples)      & 0.67 & 0.67 & 0.68 & 0.70 & \textbf{0.81} & 0.69 & 0.47 & 0.93 & 0.78 & \textbf{0.97} & 0.48 & \textbf{0.97} \\
F1 (Micro)            & \textbf{0.62} & 0.11 & 0.60 & 0.09 & 0.13 & 0.22 & 0.24 & 0.14 & \textbf{0.28} & 0.14 & 0.09 & 0.04 \\
F1 (Macro)            & 0.04 & 0.03 & 0.04 & 0.04 & 0.03 & \textbf{0.08} & 0.01 & 0.01 & 0.01 & 0.01 & 0.01 & 0.01 \\
F1 (Weighted)         & \textbf{0.62} & 0.55 & \textbf{0.62} & \textbf{0.62} & 0.58 & \textbf{0.62} & 0.19 & 0.25 & 0.24 & \textbf{0.26} & 0.22 & \textbf{0.26} \\
F1 (Samples)          & \textbf{0.64} & 0.20 & 0.62 & 0.17 & 0.22 & 0.38 & 0.24 & 0.14 & \textbf{0.28} & 0.14 & 0.09 & 0.05 \\
Hamming Loss          & \textbf{0.01} & 0.07 & \textbf{0.01} & 0.09 & 0.07 & 0.03 & \textbf{0.02} & 0.07 & 0.03 & 0.07 & 0.06 & 0.35 \\
\bottomrule
\end{tabular}
\end{table}

\subsection{Results and Discussion}
Table~\ref{tab:combined_metrics} reports micro-averaged recall, precision, and F1 for all six classifiers under both representations.
We provide detailed discussion on the results below.

\paragraph{Best precision--recall balance.} \textit{SGD} and \textit{Multinomial Naive Bayes} are the only classifiers that combine non-trivial precision with non-trivial recall, and on both representations they yield the highest F1 scores in the comparison.

With MSC codes, Multinomial Naive Bayes reaches micro-precision $0.55$, micro-recall $0.65$, and micro-F1 $0.60$, against $0.17$, $0.77$, and $0.28$ with titles; SGD reaches micro-precision $0.61$, micro-recall $0.64$, and micro-F1 $0.62$, against $0.16$, $0.47$, and $0.24$ with titles.
This is the strongest evidence in our results for a genuine precision--recall advantage of MSC codes over title embeddings, rather than a recall-only advantage: at comparable operating points, MSC-based F1 is roughly double the title-based F1.

\paragraph{High-recall, low-precision regime.} The remaining four classifiers -- \textit{Complement Naive Bayes, Passive-Aggressive, Logistic Regression}, and \textit{Neural Network} -- all trade away precision for recall, though to different degrees.

Complement Naive Bayes and Passive-Aggressive push recall highest on title embeddings (0.97 and 0.92) and to a lesser but still inflated degree on MSC codes (0.67 and 0.65), with precision collapsing to 0.05--0.08 and F1 never exceeding 0.14 on either representation; these two classifiers approach trivial near-universal labelers under our oversampling scheme, predicting membership in most of the 170 communities for nearly every paper.
Within this near-universal-labeling regime the ordering is not uniform: title embeddings post marginally higher F1 than MSC codes for Passive-Aggressive (0.14 vs.\ 0.11) and Complement NB (0.14 vs.\ 0.09).
We do not read this as title embeddings outperforming MSC codes, as both representations collapse toward near-universal labeling (precision of 0.05--0.08 for both), but it shows that MSC's advantage is not unconditional.

Logistic Regression sits at a related but distinct operating point: with MSC codes it reaches 0.79 micro-recall (against 0.48 with titles), which is the highest MSC recall of any classifier and the number we highlight for discovery-oriented use cases, but its MSC micro-precision is only 0.07 and its MSC F1 only 0.13---not meaningfully higher than the near-degenerate classifiers. 

The Neural Network behaves differently across the two representations: on titles it collapses toward the same near-universal-labeling regime as Complement NB and Passive-Aggressive (0.97 micro-recall at 0.02 micro-precision), whereas on MSC codes it lands at a more moderate 0.66 micro-recall and 0.13 micro-precision (micro-F1 0.22).
Notably, the Neural Network is the only classifier with markedly higher macro-averaged recall and F1 than the rest of the field (0.61 macro-recall and 0.08 macro-F1 with MSC codes, versus at most 0.21 macro-recall and 0.04 macro-F1 for any other classifier), suggesting it spreads predictive power across more of the 170 communities rather than concentrating on the largest ones, even though this does not translate into a higher micro-averaged score. We flag the Neural Network's per-representation gap as unstable rather than report a specific effect size: repeated training runs of this classifier show substantial epoch-to-epoch and run-to-run variability in Hamming Loss and per-class recall (observed swings of an order of magnitude in Hamming Loss across epochs of the same run), and the direction of the MSC-vs-title recall gap has itself reversed between training runs in our experiments, so the reported operating point should be read as a single representative snapshot rather than a stable, reproducible optimum; a proper paired significance test across independent retrainings is left to future work (Section~\ref{sec:limitations}).
The 0.79-recall result for Logistic Regression should therefore always be reported together with its precision; on its own it does not establish a favorable precision--recall trade-off.

Our claim is therefore narrower than ``MSC beats embeddings'': among classifiers that do not collapse into degenerate or near-degenerate labelers (SGD and Multinomial Naive Bayes), a 63-dimensional human-curated MSC vector gives a substantially better precision--recall trade-off than a 128-dimensional compressed SPECTER title embedding, roughly doubling F1.
Separately, Logistic Regression, tuned toward high recall, reaches higher recall with MSC codes than with titles (0.79 vs.\ 0.48), but this comes at low precision on both representations and should not be read as a comparably ``balanced'' result; the Neural Network instead shows the opposite pattern between representations, with its title-branch recall (0.97) exceeding its MSC-branch recall (0.66), each at correspondingly low precision, underscoring that this classifier's behavior is not consistent enough across runs to support a directional claim.
Absolute F1 scores remain modest even in the best case (peak 0.62) because three communities account for over 60\% of citations, so micro-averaged precision is dominated by the distribution head.
Macro-averaged scores, reported alongside the micro-averaged metrics in Table~\ref{tab:combined_metrics}, confirm the same broad ordering at substantially lower absolute values.

\paragraph{MSC Advantage}
We see at least four, non-exclusive, candidate explanations for why MSC codes give a better precision--recall trade-off than title embeddings in our setting.
\textbf{(1) Structural alignment}: MSC codes may directly encode mathematical domains around which software communities organize.
\textbf{(2) Discriminative power}: categorical signals partition the feature space into discrete regions, whereas embeddings may blur community boundaries.
\textbf{(3) Community formation patterns}: software may emerge from domain-specific computational needs that MSC codes capture---for example, papers on ``ML for Optimization'' and ``Neural Networks for Fluid Dynamics'' can share semantic similarity while requiring different software ecosystems, a distinction MSC codes (90-XX vs.\ 76-XX) can capture but title embeddings may miss.
\textbf{(4) Representation asymmetry}: MSC annotators have access to the full publication, whereas the title embeddings are computed from titles only (Section~\ref{sec:limitations}); part of the gap we observe may reflect this difference in input information rather than an intrinsic property of structured versus learned representations.
Our experiments cannot separate these factors from one another, and we treat them as hypotheses motivating future work rather than established causes.

\section{Conclusion}
\label{sec:conclusion}
\vspace{-2mm}
This work presents a framework for discovering and predicting mathematical software communities from scholarly data. We construct a weighted software co-usage network from publication--software relationships in the \textsf{swMATH} dataset and identify communities using the Louvain algorithm. The resulting network exhibits a strong community structure, revealing groups of software tools that are repeatedly co-cited within the scientific literature. These communities provide a complementary perspective to software mention extraction approaches by capturing higher-level relationships among tools and research domains.

We formulate publication-to-community assignment as a multi-label prediction task and investigate whether lightweight scholarly metadata representations can predict software community membership. Comparing expert-curated MSC codes with title-only SPECTER embeddings, we find that structured scholarly metadata provides a substantially stronger and more reliable signal in our setting. Across multiple classifiers, MSC-based representations achieve a more favorable precision--recall trade-off than compressed title embeddings, demonstrating that domain-specific metadata can effectively capture associations between research topics and software ecosystems. These findings do not imply that structured metadata universally outperforms semantic representations; rather, they highlight its effectiveness for large-scale software community prediction when concise, expert-curated representations are available.

Although our study focuses on mathematical software, the proposed methodology is applicable to other domains with established scholarly taxonomies, such as computing (e.g., the ACM Computing Classification System) and medicine (e.g., Medical Subject Headings). Future work will investigate hybrid representations that combine structured metadata with semantic embeddings, 
temporal evolution of software communities, and expert validation and recommendation systems based on the resulting community structure and dataset.

\section{Limitations}
\label{sec:limitations}
\vspace{-2mm}
Our approach relies on the availability of mature domain taxonomies such as MSC; domains without comparable structured classification systems may require alternative metadata sources. The current evaluation depends on existing software citations in \textsf{swMATH}, meaning that underrepresented software communities may be underestimated. Another limitation concerns the asymmetry between the two evaluated representations. MSC codes are expert-assigned using the full publication content, whereas SPECTER embeddings are generated from titles only. This design isolates representation type while ensuring consistent text availability across records, but the comparison should be interpreted as expert-curated metadata vs. title-only semantic embeddings rather than a fully matched comparison of information content. Additionally, community assignments are derived from software co-usage patterns and have not yet been independently validated. The Neural Network results exhibit higher training variability than the other models and should therefore be interpreted with caution.

\begin{credits}

\subsubsection{\ackname}
This work was supported by the DFG project ``\textit{Enhancing the discoverability of mathematical source code}'' (Grant No.~\href{https://gepris.dfg.de/gepris/projekt/561181416}{561181416}) and the EU Horizon Europe \textit{LUMEN} project (Grant No.~\href{https://cordis.europa.eu/project/id/101187940}{101187940}). Views and opinions expressed are those of the authors only and do not necessarily reflect those of the funding organizations or the granting authority.

\end{credits}

\bibliographystyle{splncs04}
\bibliography{references}
\end{document}